\documentclass[twocolumn]{aastex62}

\usepackage{graphicx} 
\usepackage{array, geometry}
\usepackage{threeparttablex, booktabs}
\usepackage{natbib}
\usepackage{amsmath}
\usepackage{xcolor}
\usepackage{url}

\submitjournal{PASP}
\shorttitle{Multipole analysis}
\shortauthors{Leahy, Ranasinghe, Filipovic \& Smeaton}

\begin{document}

\title{Multipole Functions for Image Analysis II: Equal Area Weighting and Application to Supernova Remnant Images}

\correspondingauthor{Denis Leahy}
\email{leahy@ucalgary.ca}

\author{D.A. Leahy}
\affiliation{Department of Physics and Astronomy, University of Calgary, Calgary, AB T2N 1N4, Canada}

\author{S. Ranasinghe}
\affiliation{Department of Physics and Astronomy, University of Calgary, Calgary, AB T2N 1N4, Canada}

\author{M. D. Filipovi\'c}
\affiliation{Western Sydney University, Locked Bag 1797, Penrith South DC, NSW 2751, Australia}

\author{Z. J. Smeaton}
\affiliation{Western Sydney University, Locked Bag 1797, Penrith South DC, NSW 2751, Australia}
\date{Jnauary 2025}

\begin{abstract}

New basis functions for 2 dimensional (2D) image analysis with a circular boundary (referred to as multipole analysis) are derived which 
are equal-area weighted. 
We present open access Python code hosted by GitHub,  with which users can apply the multipole analysis to images.
The new multipole analysis is applied to a set of 28 supernova remnants (SNRs) which are selected to have both radio and X-ray images, and have been identified as Type Ia or Type CC. 
Each pair of SNR images (radio and X-ray) was convolved to the same spatial resolution prior to analysis. 
The resulting multipole radial powers and angular powers, from order 0 to 5, for a given SNR are different for different multipoles and for a given multipole are different between X-ray and radio images.
The X-ray radial powers (for orders $>$0) are larger on average than the radio radial powers (more radial structure in X-rays than radio). 
The angular powers are smaller than the radial powers on average (more radial structure than angular structure).  
Comparing Type Ia and Type CC populations, the radial powers (for orders $>$0) are on average larger for Type CC than Type Ia for X-ray and radio images, with larger difference for X-ray images. 
The angular powers (for orders $>$0) are similar between Type Ia and Type CC for both radio and X-ray images.   

\end{abstract}


\section{Introduction}\label{sec:intro}

\subsection{Image Analysis}

Quantitative image analysis of astronomical objects can reveal properties of the structure of the object under study.
For many objects of interest, the emission is optically thin, which implies that the image is the projection of the 3-dimensional (3D) emissivity distribution onto the plane of the sky. 
Because the observations (the image) are 2D, the analysis of the image is also 2D.  

The simplest 3D structure is uniform emission in a sphere which yields a 2D image with circular symmetry and radially decreasing surface brightness. Deviations from 3D spherical symmetry yield more complex 2D images, which generally exhibit both radial and angular dependencies, e.g. as expressed in spherical polar coordinates. 

For 3D structures in spherical polar coordinates (with a spherical boundary) the spherical harmonics form suitable basis functions used for analysis. 
They form a complete set of orthogonal basis functions with symmetry appropriate to the coordinates.

However, for 2D images the suitable basis functions are not as familiar.
The current work is a followup study of our first study \citep{2025PASP..137f4502L}, which corrected important errors in the astronomical literature on multipole image analysis.
That work derived appropriate orthogonal basis functions for analysis of 2D images with a circular boundary. 
Surprisingly, multipole analysis of 2D astronomical images had not been addressed correctly, or in any depth, previously in the astronomical literature. 
The current work presents an important extension to that of \cite{2025PASP..137f4502L}: the basis functions for equal-area weighting are derived, then applied. 
Although more complex, equal area weighted basis functions are much more natural compared to the simplest set of unweighted orthogonal basis functions previously presented. 

\subsection{Application to Supernova Remnants (SNRs)}

A SNR is optically thin at radio and X-ray wavelengths, thus the radio or X-ray image is a projection of the 3D volume emission onto the plane of the sky.
The analysis of the 2D image of a SNR measures the degree of non-uniformity of the sky projection of the SNR, which is a measure of the non-uniformity of the 3D SNR structure.

After deriving the basis functions, we present the analysis of a set of SNR images observed at radio and X-ray wavelengths. 
Section 2 summarizes the  mathematics of the expansion of 2D images with a circular boundary in terms of orthogonal basis functions and derives the basis functions for the case of equal-area weighting. 
Section 3 describes the set of SNRs for which there are both X-ray and radio images. 
The results of the multipole analysis are presented in  Section 4. 
In Section 5, we discuss the results of the multipole analysis applied to SNRs. 
The work is summarized in Section 6.

\section{Methods}\label{sec:meth}

\subsection{Image Representation by a Sum of Orthogonal Multipole Functions}\label{complete}

Complete sets of functions are derived from solutions of self-adjoint differential equations \citep{mathphys2,mathphys1}. 
This discussion is a summarized version of the more complete description given in Section 2 of \cite{2025PASP..137f4502L}, with some additions such as uniform area weighting.

To completely represent an image (a 2D function) one requires a complete set of 2D functions with the same boundaries as the image. 
For example, a rectangular image can be represented in terms of Fourier series in rectangular (x,y) coordinates.
However, many astronomical images have a basic circular symmetry, with small or large perturbations, so are better represented in 2D plane polar coordinates with a circular boundary.
For the latter case, the first requirement is to find a complete set of orthogonal functions in 2D polar coordinates $0<\rho<R$, $0<\phi<2\pi$, where $R$ is the radial coordinate of the boundary. 
By rescaling the outer boundary to unity, one has $r=\rho/R$, $0<r<1$.

The chosen self-adjoint differential operator in plane polar coordinates is $\nabla^2 \Phi(r,\phi)=0$, with boundary conditions $\Phi(r,\phi)=\Phi(r,\phi+2\pi)$ (i.e. single valued); and $\Phi(0,\phi)$ and $\Phi(1,\phi)$ are finite. 
The complete set of functions consists of products of radial functions $r^n$, $n=0,1,2...\infty$, and angular functions. 
The angular functions are 
$cos(m\phi)$, $m=0,1,2...\infty$ and $sin(m\phi)$, $m=1,2...\infty$. 
$m$ is the angular order of the multipole.

The radial functions $r^n$, $n=0,1,2...\infty$, are not orthogonal.
We apply Gramm-Schmidt ortho-normalization to $r^n$ over the interval $0<r<1$ with the weight function $w(r)=r$.
The reason for the weight function is that one desires to have multipole coefficients which are uniformly weighted over the area of an astronomical image, and the area element $dA$ in polar coordinates is given by $dA=dr~rd\phi$.

The complete and orthonormal basis functions are:
\begin{equation}
\label{eqn.2}
fc_{n,0}(r,\phi)=P_{n}^{r}(r)
\end{equation}
$n=0,1,2...\infty$, m=0; 
\begin{equation}
\label{eqn.3}
fc_{n,m}(r,\phi)=P_{n}^{r}(r) ~cos(m\phi)
\end{equation}
$n=0,1,2...\infty$, $m=1,2...\infty$; and 
\begin{equation}
\label{eqn.4}
fs_{n,m}(r,\phi)=P_{n}^{r}(r) ~sin(m\phi)
\end{equation}
$n=0,1,2...\infty, m=1,2...\infty$,
where we denote the radial functions with weight function $w(r)=r)$ as $P_{n}^{r}(r)$.
Above $n$ is known as the radial order of the multipole. 

\begin{figure*}
    \centering
    \includegraphics[width=1\linewidth]{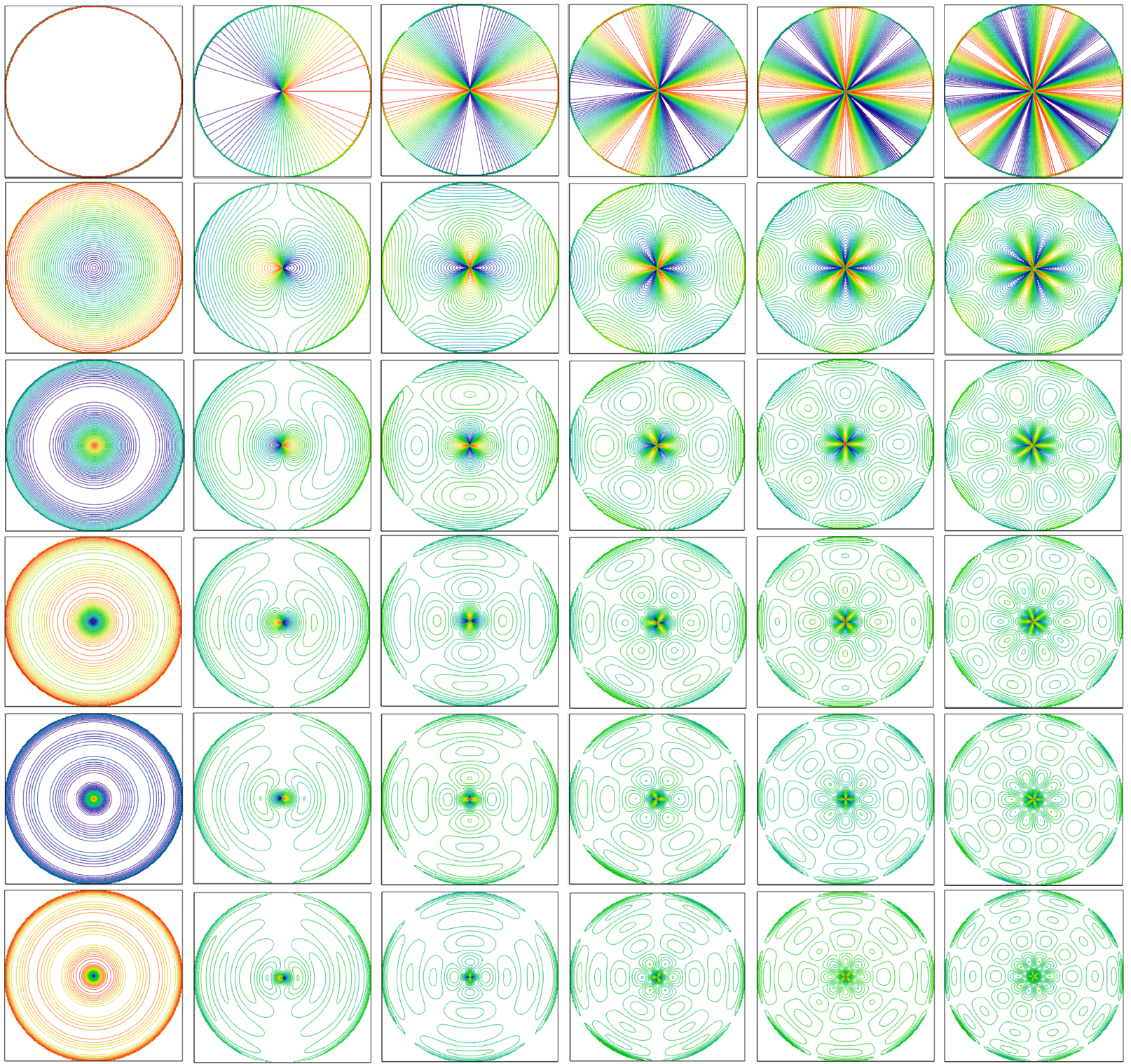} 
    \caption{The basis functions $fc_{n,m}(r,\phi)$ for $n$=0, 1, 2, 3, 4 and 5 (rows 1 to 6)
    and for $m$=0, 1, 2, 3, 4 and 5 (columns 1 to 6). 
    The color scale ranges from red (most positive) through to purple (most negative), as seen in linear radial scale for $fc_{1,0}(r,\phi)$ (first column, second row).
    The $fs_{n,m}(r,\phi)$ basis functions (for $m\ne$0) are the same except that they are rotated in angle counterclockwise by 90$^\circ/m$.}
    \label{fig:basis}
\end{figure*}

We give the radial functions for $n=0$, 1, 2, 3, 4 and 5 here.
\small
\footnotesize
\begin{eqnarray}
 P_{0}^{r} &=& \sqrt{2} \nonumber \\  
P_{1}^{r} &=& 6~(r-\frac{2}{3}) \nonumber \\ 
P_{2}^{r} &=& 10~\sqrt{6}~(r^2-\frac{6}{5}r+\frac{3}{10}) \nonumber \\ 
P_{3}^{r} &=& 70~\sqrt{2}~(r^3-\frac{12}{7}r^2+\frac{6}{7}r-\frac{4}{35}) \nonumber \\ 
P_{4}^{r} &=& 126~\sqrt{10}~(r^4-\frac{20}{9}r^3+\frac{5}{3}r^2-\frac{10}{21}r+\frac{5}{126}) \nonumber \\ 
P_{5}^{r} &=& 924~\sqrt{3}~(r^5-\frac{30}{11}r^4
+\frac{30}{11}r^3-\frac{40}{33}r^2+\frac{5}{22}r-\frac{1}{77}) \nonumber\\ 
\end{eqnarray}
\normalsize
The cosine basis functions are illustrated in Figure~\ref{fig:basis} for $n,m$ =0, 1, 2, 3, 4 and 5. 
The sine basis functions are the same but rotated by 90$^\circ$. 
The multipole expansion of the image $(F(r,\phi))$ in terms of the basis functions and coefficients, $ac_{n,m}$ and $as_{n,m}$, is the same as given in Section 2.2 of \cite{2025PASP..137f4502L}.

The multipole coefficients, $ac_{n,m}$ and $as_{n,m}$, can be summarized by the multipole power ratios 
(hereafter called powers).
For radial order $n$ and angular order $m$ the power $P(n,m)$ is given by:
\begin{equation}
\label{eqn.8}
P_{n,m}=(ac_{n,m}^2+as_{n,m}^2)/a_{0,0}^2   
\end{equation}
where $as_{n,0}$ is defined as $0$ to avoid having a separate definition for the case $m=0$ for the sine basis functions. 
These are normalized to be independent of the total flux of the image by dividing by $a_{0,0}^2$. 
The powers are a measure of the brightness in an image in each radial and angular order.  
However, to fully represent the image the full set of coefficients $ac_{n,m}$ and $as_{n,m}$ are required.

Radial multipole powers which are summed over all angular orders $m$ are given by:
 \begin{equation} \label{eqn.9}
P_{rad,n}=\sum_{m=0}^{\infty}(ac_{n,m}^2+as_{n,m}^2)/a_{0,0}^2   
\end{equation}
and angular powers which are summed over radial orders $n$ are given by:
 \begin{equation} \label{eqn.10}
P_{ang,m}=\sum_{n=0}^{\infty}(ac_{n,m}^2+as_{n,m}^2)/a_{0,0}^2   
\end{equation}

In practise, only a finite number of radial and angular modes are used for image analysis. 
E.g., one may only be interested in the coarse structure of an image.
Modes which have finer scale than the image resolution are not meaningful, and 
small scale (large $n$ and $m$) may be  dominated by noise. 

\subsection{Python based software for calculating multipole coefficients and powers}\label{software}

\begin{figure*}
    \centering
    \includegraphics[width=\linewidth]{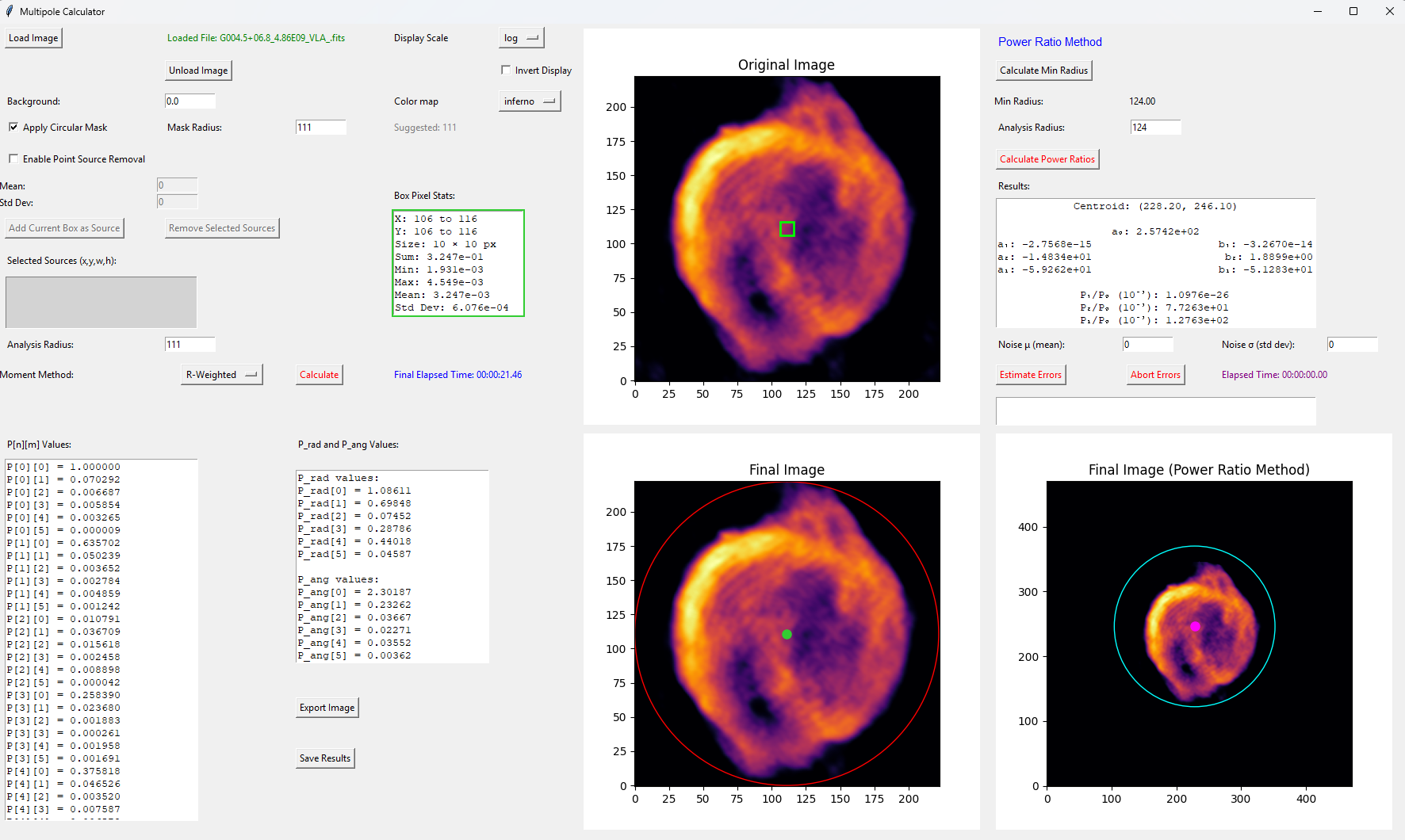}   
    \caption{Screenshot of the multipole calculator graphical user interface. The upper-left panel provides image loading and parameter input controls, while the lower-left panel displays the radial and angular power results along with options to export results and   images. The central panels show the input image and image with center and boundary. The upper right panel has input parameters for the power method, optional error estimation, with  the lower-right image showing the boundary and centroid for the PRM method. }
    \label{fig:Screenshot}
\end{figure*}

We present a Python-based graphical user interface (GUI) developed for the quantitative analysis of astronomical images using multipole moment techniques
(\url{https://doi.org/10.5281/zenodo.19013723}).
The software is designed to facilitate the computation of radial and angular powers from two-dimensional intensity distributions, such as X-ray or radio observations stored in FITS format. 
Emphasis is placed on usability and reproducibility, enabling users to perform image-based morphological analyses in a consistent manner.
The README.txt file outlines the basic functionality of the software and its dependencies. 

Figure \ref{fig:Screenshot} shows a screenshot of the GUI. The software supports FITS files, allowing users to load and analyze astronomical images directly. 
It is recommended that the FITS file be cropped such that the full extent of the source lies within the image frame. 

Once an image is loaded, users may apply a background level, which is subtracted uniformly from all pixels. 
A circular mask with a user-specified radius can then be applied. By default, the mask radius is set to extend from the image center to the edge, with all pixel values outside this radius set to zero.

An additional option allows for the removal of point sources from the image. 
This is accomplished by positioning the green selection box over a nearby background region to compute the local pixel statistics, which are displayed in the Box Pixel Stats panel adjacent to the original image. 
After specifying the mean and standard deviation derived from this region, the green box can be placed over the point source, and the enclosed pixels are replaced with values drawn from a normal distribution, effectively removing the source while preserving the surrounding noise characteristics.

For the multipole moment analysis, three calculation methods are available, with the default set to the r-weighted formulation described in Section~\ref{sec:meth}. 
The first option computes power coefficients for orders $n$ and $m$ up to 3, following the approach of \cite{2025PASP..137f4502L}. 
The second option employs the same formalism but extends the calculation to $n,m \leq 5$. Once the Calculate button is pressed, the resulting power coefficients are displayed in the lower-left panels. 
At this stage, any modifications to the image are reflected in the lower-middle panel, which also shows the effective calculation region and center. 
Results may be exported as a CSV file, and the original image can also be saved.

The upper-right panel provides access to the power-ratio method (PRM) as described by \cite{2011ApJ...732..114L,2019arXiv190911803R}. 
Because centroid determination in this method can result in an effective radius extending beyond the original image boundaries, the image is automatically padded and a new effective radius must be specified before computation. Error estimates are calculated following the methodology of \cite{2019arXiv190911803R}. 
The lower-center image shows the circular boundary and image center of the SNR (used in the current work); the lower-right image shows an alternate circular boundary and centroid for the PRM method.

\section{Supernova Remnant X-ray and Radio Image Sample}\label{sec:sample}

\begin{figure*}
    \centering
 \includegraphics[width=\linewidth]{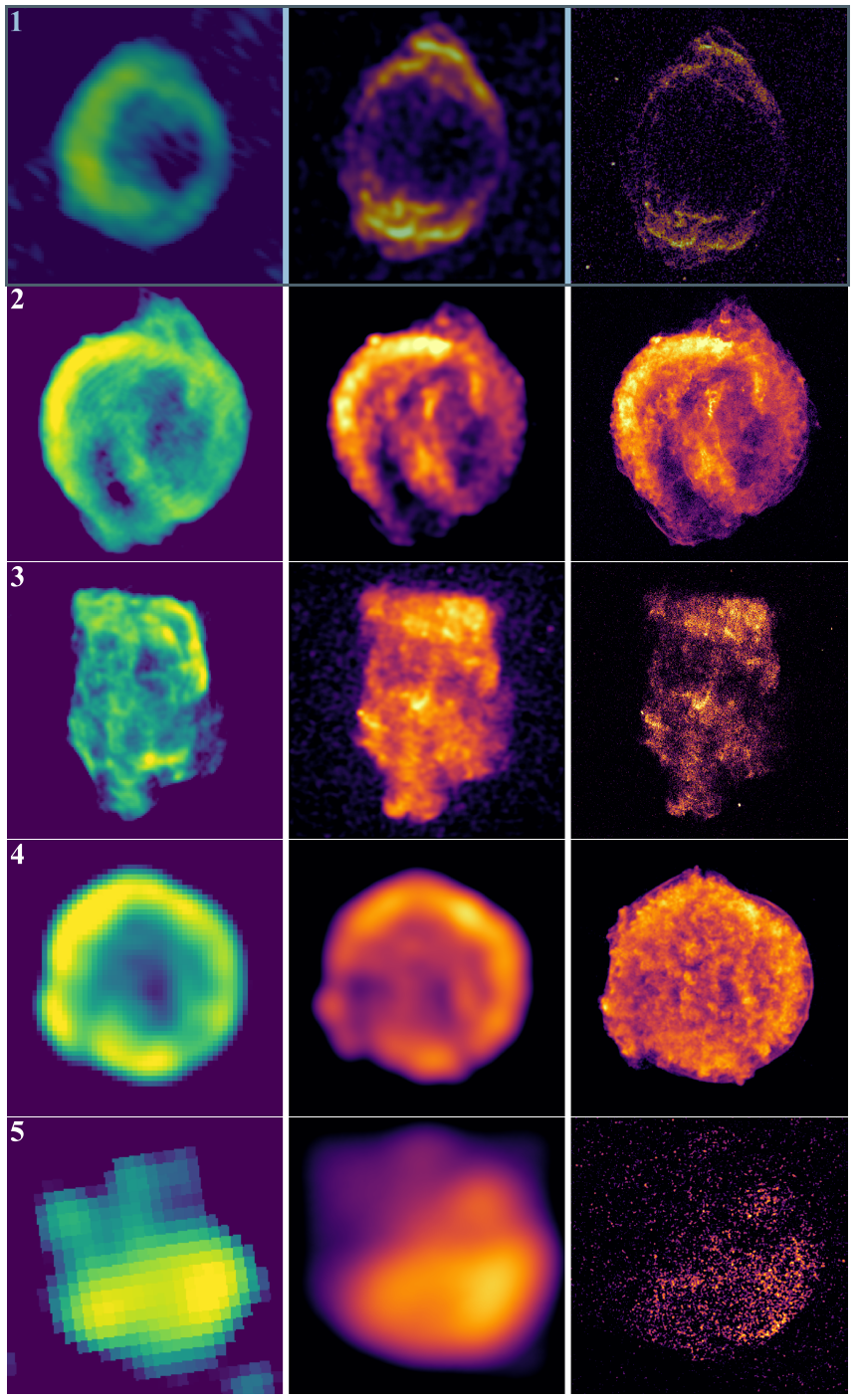}
    \caption{The images for SNRs 1 to 5 listed in Table~\ref{tab1}, with the radio image at left, the smoothed X-ray image at center and the unsmoothed X-ray image at right. Panels 1–11 are of Type Ia SNRs, while panels 12–28 are of core-collapse SNRs.}
    \label{fig:images1}
\end{figure*}

\begin{figure*}
    \centering
\includegraphics[width=1\linewidth]{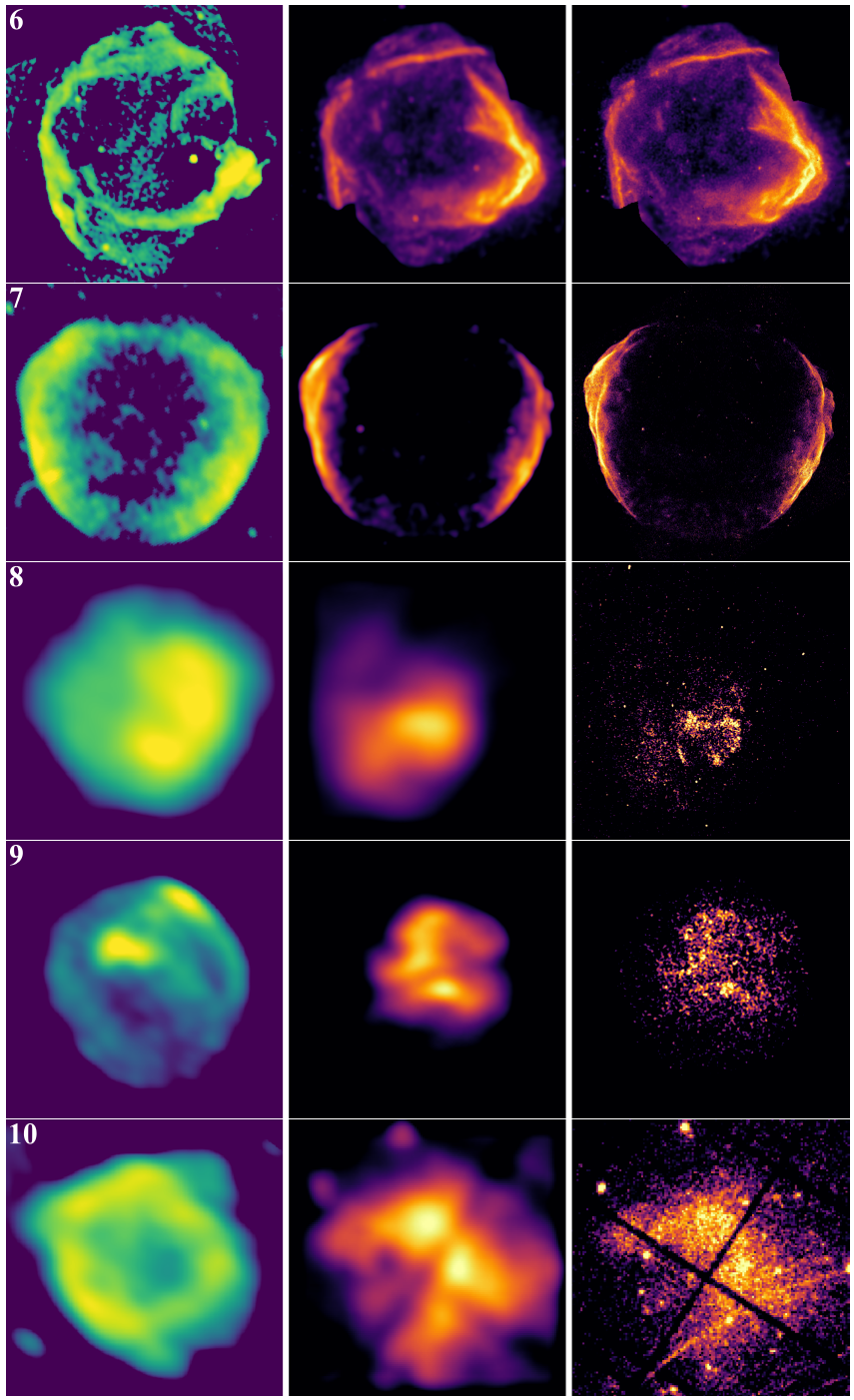}
    \caption{The images for SNRs 6 to 10 listed in Table~\ref{tab1}, with the radio image at left, the smoothed X-ray image at center and the unsmoothed X-ray image at right.}
    \label{fig:images2}
\end{figure*}

\begin{figure*}
    \centering
\includegraphics[width=1\linewidth]{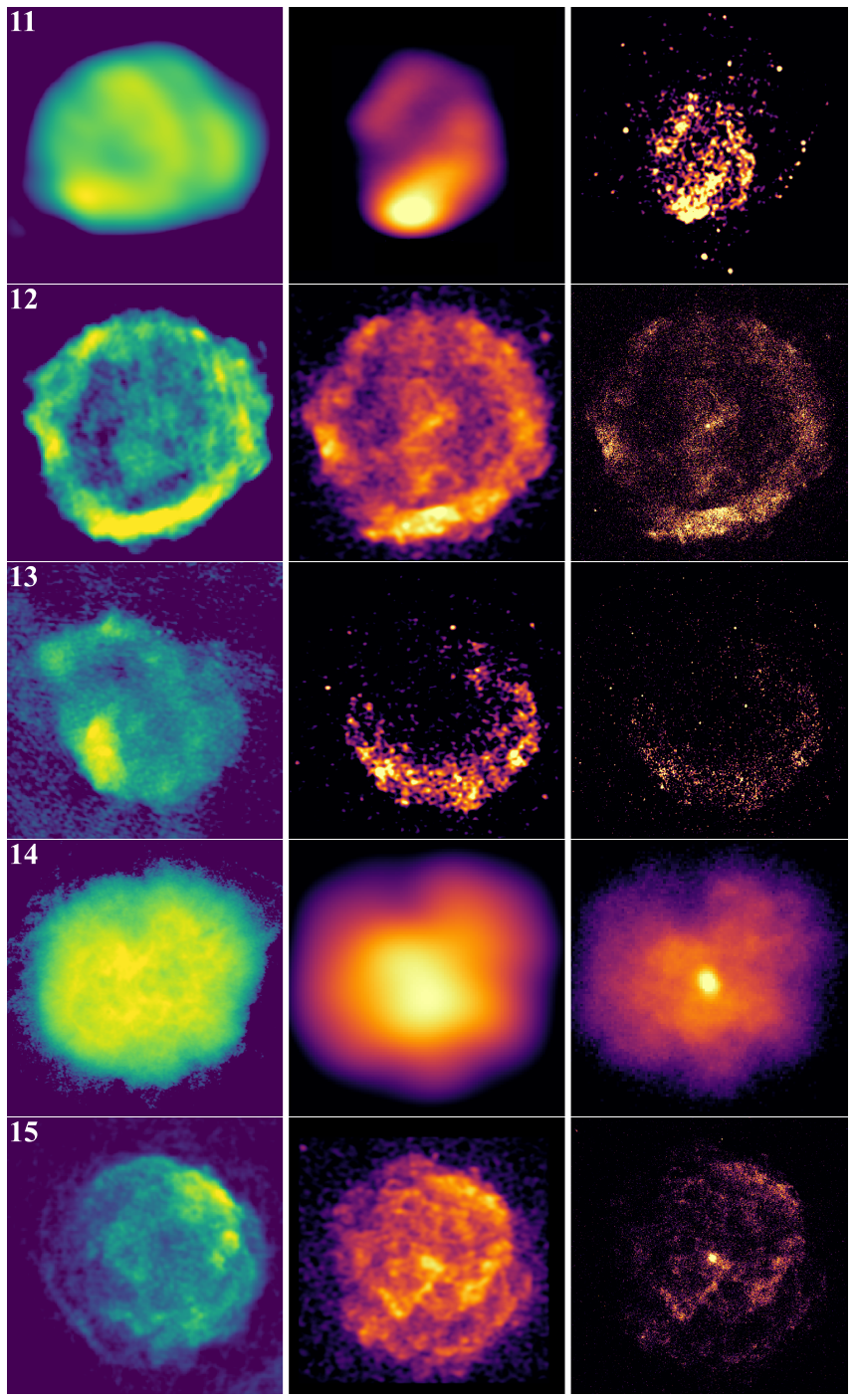}
    \caption{The images for SNRs 11 to 15 listed in Table~\ref{tab1}, with the radio image at left, the smoothed X-ray image at center and the unsmoothed X-ray image at right.}
    \label{fig:images3}
\end{figure*}

\begin{figure*}
    \centering
\includegraphics[width=1\linewidth]{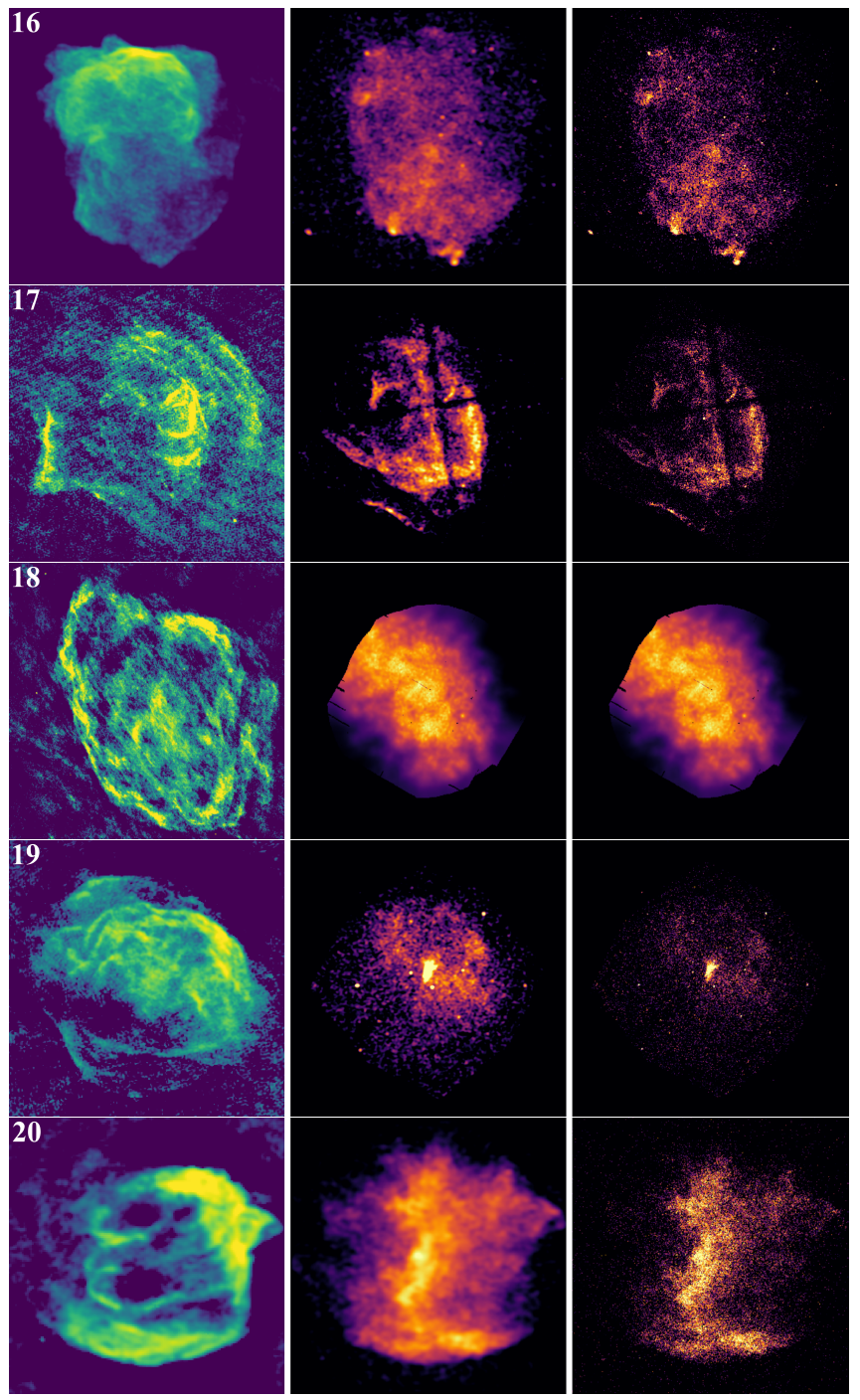}
    \caption{The images for SNRs 16 to 20 listed in Table~\ref{tab1}, with the radio image at left, the smoothed X-ray image at center and the unsmoothed X-ray image at right.}
    \label{fig:images5}
\end{figure*}

\begin{figure*}
    \centering
\includegraphics[width=1\linewidth]{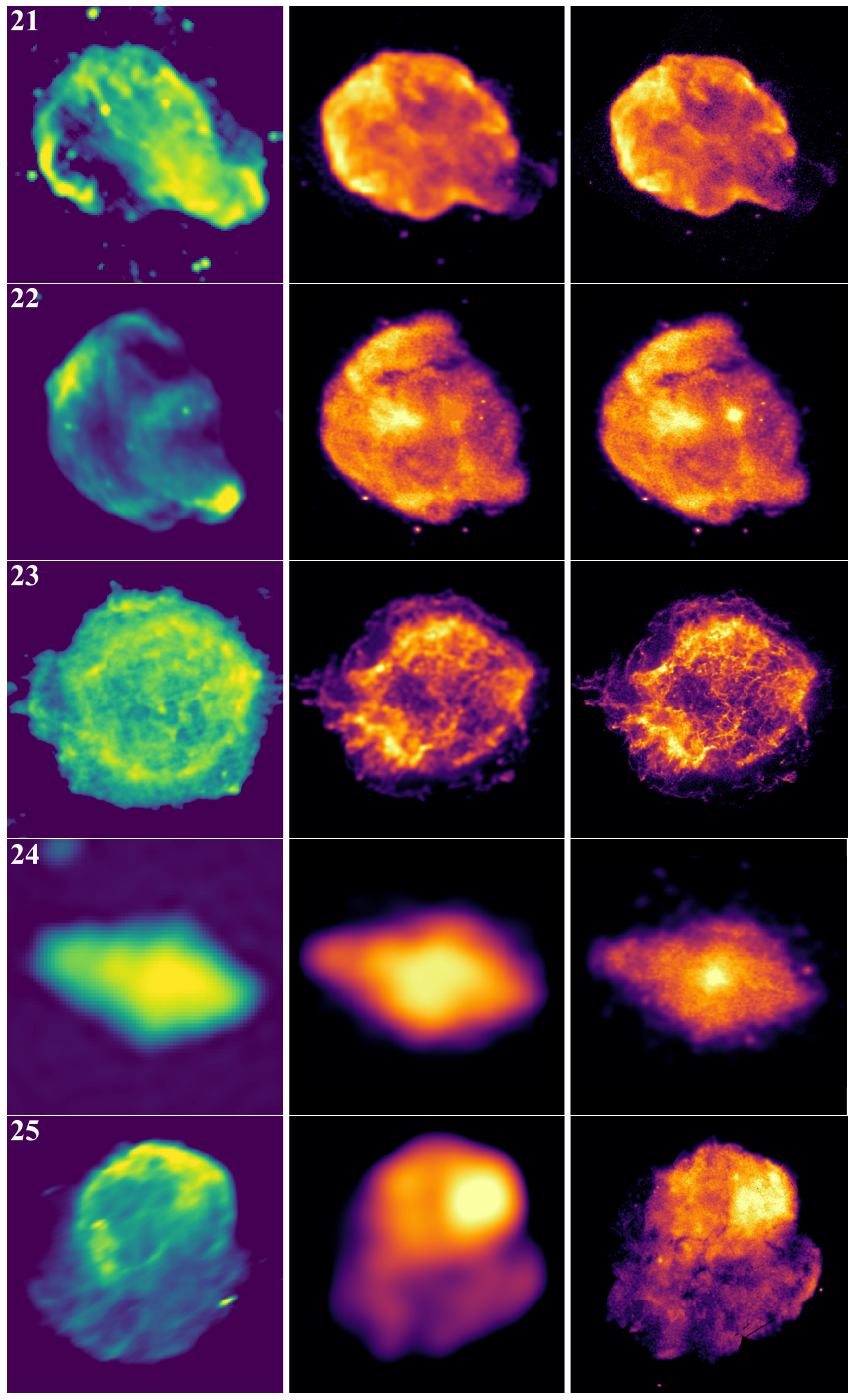}
    \caption{The images for SNRs 21 to 25 listed in Table~\ref{tab1}, with the radio image at left, the smoothed X-ray image at center and the unsmoothed X-ray image at right.}
    \label{fig:images6}
\end{figure*}

\begin{figure*}
    \centering
\includegraphics[width=1\linewidth]{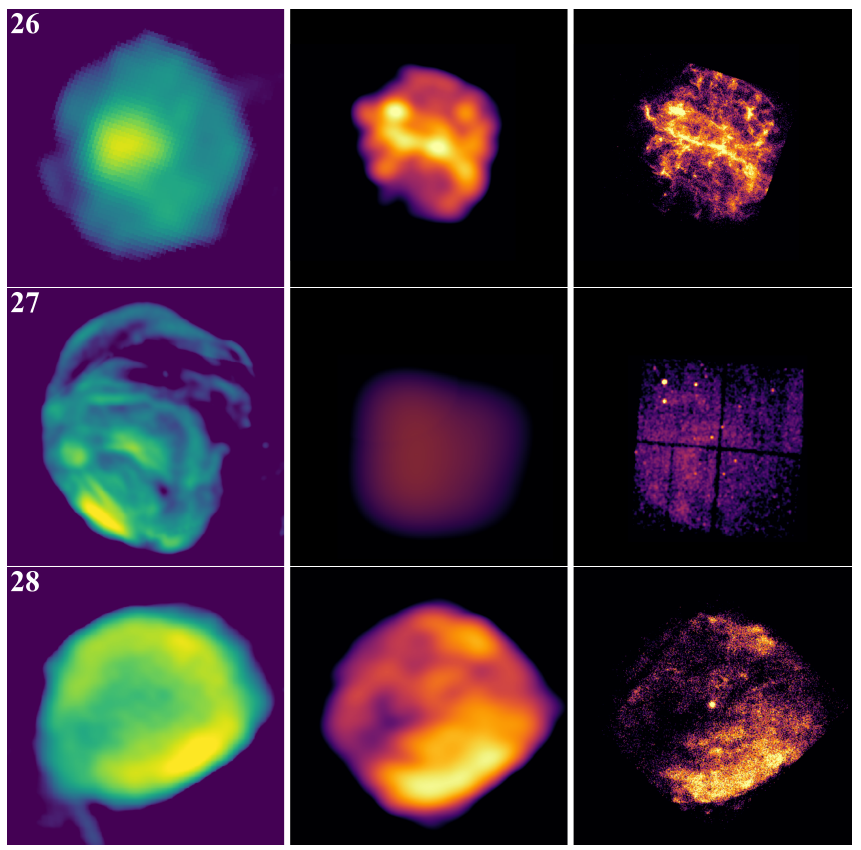}
    \caption{The images for SNRs 26 to 28 listed in Table~\ref{tab1}, with the radio image at left, the smoothed X-ray image at center and the unsmoothed X-ray image at right.}
    \label{fig:images7}
\end{figure*}

\begin{table}[!ht]
    \centering
    \caption{Supernova Remnant Sample}\label{tab1}
    \begin{tabular}{llcc}
    \hline
    \# $\quad$  & SNR      & Type$^\textrm{a}$  &  Radio Size$^\textrm{b}$           \\
        &                   &       &  (arcmin)             \\
    \hline 
    \multicolumn{4}{l}{Type Ia}             \\
	\hline								
	1	&	G$1.9	+0.3$	&	S	&	$1.5 \times 1.5$	\\
	2	&	G$4.5	+6.8$	&	S	&	$3 \times 3$	    \\
	3	&	G$41.1	-0.3$	&	S	&	$4.5 \times 2.5$	\\
	4	&	G$120.1	+1.4$	&	S	&	$8 \times 8$	    \\
	5	&	G$306.3	-0.9$	&	S?	&	$4 \times 4$	    \\
	6	&	G$315.4	-2.3$	&	S	&	$42 \times 42$	    \\
	7	&	G$327.6	+14.6$	&	S	&	$30 \times 30$	    \\
	8	&	G$337.2	-0.7$	&	S	&	$6 \times 6$	    \\
	9	&	G$344.7	-0.1$	&	C?	&	$8 \times 8$	    \\
	10	&	G$346.6	-0.2$	&	S	&	$8 \times 8$	    \\
	11	&	G$352.7	-0.1$	&	S	&	$8 \times 6$        \\
	\hline									
	\multicolumn{4}{l}{Core-Collapse} \\	
	\hline	
	12	&	G$11.2	-0.3$	&	C	&	$4 \times 4$       	\\
	13	&	G$15.9	+0.2$	&	S?	&	$7 \times 5$      	\\
	14	&	G$21.5	-0.9$	&	C	&	$5 \times 5$       	\\
	15	&	G$27.4	+0.0$	&	S	&	$4 \times 4$       	\\
	16	&	G$31.9	+0.0$	&	S	&	$7 \times 5$       	\\
	17	&	G$33.6	+0.1$	&	S	&	$10	\times 10$     	\\
	18	&	G$34.7	-0.4$	&	C	&	$35 \times 27$     	\\
	19	&	G$39.2	-0.3$	&	C	&	$8 \times 6$       	\\
	20	&	G$43.3	-0.2$	&	S	&	$4 \times 3$       	\\
	21	&	G$74.0	-8.5$	&	S	&	$230 \times	160$	\\
	22	&	G$109.1	-1.0$	&	S	&	$28 \times 28$	    \\
	23	&	G$111.7	-2.1$	&	S	&	$5 \times 5$      	\\
	24	&	G$130.7	+3.1$	&	F	&	$9 \times 5$      	\\
	25	&	G$189.1	+3.0$	&	C	&	$45 \times 45$    	\\
	26	&	G$292.0	+1.8$	&	C	&	$12 \times 8$     	\\
	27	&	G$327.4	+0.4$	&	S	&	$21 \times 21$    	\\
	28	&	G$332.4	-0.4$	&	S	&	$10 \times 10$    	\\
	\hline								
    \end{tabular}
\tablecomments{\\$^{\textrm{a}}$: SNR types: S: shell, F: filled center, and C: composite.\\
$^{\textrm{b}}$: Major and Minor Axes}
\end{table}

The current analysis compares X-ray and radio structures of SNRs, and also compares the structures of Type Ia and core-collapse (CC) SNRs.
Thus, our sample was selected from sources that possess both radio and X-ray observations, exhibit comparable spatial extent, are relatively bright, and have confirmed classifications as either Type Ia or CC. 

To minimize uncertainties related to surface brightness and spatial resolution that can affect morphological measurements, our sample is restricted to SNRs within the Milky Way. 
Based on these criteria, we selected 11 Type Ia and 17 CC SNRs for analysis. The full sample is presented in Table~\ref{tab1}, along with the shell type and radio sizes as presented by \cite{2025JApA...46...14G}.

Images for both X-ray and radio were obtained from the SNR Catalog (SNRcat; \url{http://snrcat.physics.umanitoba.ca}, \cite{2012AdSpR..49.1313F}). 
The SNR Catalog lists the observatory (most often Chandra, Suzaku, or XMM-Newton) used for each X-ray observation as well as the list of references for the X-ray observation.
Green's SNR Catalog (\url{https://www.mrao.cam.ac.uk/surveys/snrs/},  \cite{2025JApA...46...14G}) has the reference list for the radio observations (most often VLA, ATCA or MOST) of each SNR.  

\begin{table*}[h!]
\begin{center}
\caption{Effect of Smoothing of X-ray Images on Powers} 
      \centering
    \begin{tabular}{lccccccccc}
 \hline
 X-ray Radial power \\
    $n$ value: & 0 & 1  & 2 & 3 & 4& 5 \\
   \hline
unsmoothed Mean & 1.141 &	1.096&	0.787&	0.764&	0.575&	0.430 \\ 
smoothed Mean & 1.130 &	1.064&	0.701&	0.671&	0.504&	0.376 \\
decrease  & 0.011 & 0.032 &	0.086&	0.093&	0.071&	0.054 \\ 
Uncertainty$^{\textrm{b}}$ & 0.02 & 0.02 & 0.02 & 0.02 & 0.02 & 0.02\\
unsmoothed SD$^{\textrm{a}}$ & 0.165  &	0.795&	0.976&	0.875&	0.761& 0.610 \\
smoothed SD & 0.162  &	0.655&	0.676&	0.643&	0.622& 0.529 \\
\hline
X-ray Angular power \\
    $m$ value: & 0 & 1  & 2 & 3 & 4& 5 \\
   \hline
unsmoothed Mean & 4.220	&	0.282	&	0.132	&	0.056	&	0.074	&	0.029  \\ 
smoothed Mean & 4.034	&	0.223	&	0.103	&	0.036	&	0.035	&	0.016  \\ 
decrease       & 0.186 &    0.059   &	0.029  &    0.020 &	    0.039  &   0.013 \\ 
unsmoothed SD & 3.506     &	0.298&	0.144&	0.061 &	0.123&	0.035  \\
smoothed SD & 2.594     &	0.285&	0.143&	0.058 &	0.046&	0.034  \\
\hline
\hline
    \end{tabular}
    \label{tab:origVsm}
\end{center}
\tablecomments{$^{\textrm{a}}$: SD means standard deviation.}
\tablecomments{$^{\textrm{b}}$: The calculation of uncertainty in individual radial and angular powers is described in Section~\ref{sec:SNRmult}. They are the same for different multipoles, thus this row is not repeated in the table for the different cases in Column 1.}
\end{table*}

The extracted data were cropped using CCDLAB \citep{2025PASP..137f4502L} such that the emission was entirely contained within the image frame. 
For each pair (radio and X-ray) of images of a given SNR, the higher resolution image of the two was convolved to the resolution of the lower resolution image.
For all 28 SNRs, the X-ray image had higher resolution, so it was smoothed to the radio resolution using an elliptical Gaussian with major and minor axes ($\sigma_{maj}$ and $\sigma_{min}$) given by:
\begin{eqnarray}
    \sigma_{maj}&=&\sqrt{\sigma_{maj,radio}^2-\sigma_{maj,xray}^2} \nonumber \\
    \sigma_{min}&=&\sqrt{\sigma_{min,radio}^2-\sigma_{min,xray}^2}
\end{eqnarray}
Thus for comparison of radio vs. X-ray images, the resolutions of the 
radio and convolved X-ray images are the same.
In Figure~\ref{fig:images1} to Figure~\ref{fig:images7}  we show the radio image (left) and original (right) and convolved (center) X-ray images for each of the 28 SNRs.

Table~\ref{tab:origVsm} shows the effects of smoothing of the X-ray images on the multipole powers.
The individual multipole powers are decreased slightly in all cases by smoothing, which results in a small decrease in the mean radial and angular powers, as shown in Table~\ref{tab:origVsm}. 
The decrease in radial powers is on average 0.058 and decrease in angular powers (excepting $m=0$) is on average 0.032.
The decreases in powers results in the standard deviations to decrease, in most cases by small amounts.
The large decrease in the angular $m=0$ standard deviation is caused by a single SNR ( G130.7+0.3) which a change in $m=0$ angular power from 17.44 (before smoothing) to 8.92 (after smoothing).

\section{Multipoles for the SNR Images}\label{sec:SNRmult}

We chose a circular boundary for each image as the minimum-sized circle which encompasses the emission from the SNR in both radio and X-ray bands. Thus 
we used the radio image to define the boundary because the radio emission extended outward as far or further than the X-ray emission
for all SNRs in the sample. 
The center of that circle is here called the ``image center" of the SNR\footnote{
Using the centroid as center of analysis yields a radius larger than using the image centre, and it necessarily includes more area with no SNR emission inside the larger radius.
A comparison of the positions of centroid with image center is given by \cite{2025PASP..137f4502L} for 24 SNR X-ray images (their Fig.2): in about half of the cases there is a significant offset of the centroid from the image center.}.

The radius values were scaled so that $r=1$ at the circle boundary of each SNR. 
The analysis utilized the modes: $n$= 0, 1, 2, 3, 4 and 5 and $m$= 0, 1, 2, 3, 4 and 5. 
We calculated the coefficients, $ac_{n,m}$ and $as_{n,m}$, and from them determined the multipole powers, then the radial and angular powers for the SNRs listed in Table \ref{tab1}. 
I.e., 36 X-ray and 36 radio multipole powers  were calculated for each of the 28 SNRs, then the 6 X-ray and 6 radio radial and 6 angular X-ray and 6 radio multipole powers were determined from those.

The errors in the multipole powers were discussed by \cite{2025PASP..137f4502L}.
That work found that the errors in the calculated multipole powers are typically $\simeq$0.02 or smaller.
The uncertainties in multipole powers were estimated as follows.
Noise was added to each image pixel using a Gaussian random number generator with zero mean and standard deviation based on the background values for that SNR. 
This was done to create multiple versions ($\sim$20) of images of each SNR with different noise realizations, which were used to calculate multiple versions of the multipole powers. 
The standard deviations of these versions were taken as the contribution of noise to each multipole power. 
We also calculated the change in multipole powers caused by an offset of one pixel in different directions of the image center. 
The latter contribution dominated over the background contribution by a factor of $\sim$5.
The above was done for a subset of the SNRs for different radial and angular powers, 
yielding a maximum uncertainty of $\simeq$0.02 and mean of $\simeq$0.005. 
There is no variation between different multipole powers (different radial $n$ or angular $m$ values) because the added noise had no position dependence.
We took the noise (uncertainty) value as 0.02.
The errors in the multipole powers from noise are small compared to the difference in multipole powers between SNRs.

Because three of the X-ray images have artifacts caused by the detector gaps in the Chandra ACIS (SNR numbers 10, 17 and 27), we also calculated the multipole power statistics excluding those three SNRs. 
Comparing the original sample of 28 SNRs to the sample of 25 SNRs (after excluding the ones with artifacts), we find that the average radial and angular powers for the radio images change by 0.001 to 0.03, with mean change of 0.018. 
The average radial and angular powers for the X-ray images change by 0.001 to 0.253, with mean change of 0.059. 
The X-ray power that changes the most is $P_{ang,0}$, and excluding that on the average change in X-ray power reduces to 0.030. 
For both radio images and X-ray images the radial and angular powers are slightly larger on average when excluding the SNRs with artifacts (numbers 10 17 and 27), because SNRs 17 (G346.6-0.2)  and 27 (G327.4+0.4) are slightly smoother on average than the other SNRs in the sample.
The artifacts in the X-ray images thus do not seem to have any significant effect on the multipole powers.
Thus the values below are presented for the full 28 SNR sample.

\subsection{Comparison of Powers Between Radio and X-ray Images of SNRs}

\begin{table*}[h!]
\begin{center}
\caption{Statistics of Radio and X-ray Powers and Their Differences (X-ray minus Radio)} 
      \centering
    \begin{tabular}{lccccccccc}
 \hline
  Radial power (radio) \\
    $n$ value: & 0 & 1  & 2 & 3 & 4& 5 \\
   \hline
Mean & 1.166	&	0.977	&	0.520	&	0.471	&	0.292	&	0.185 \\
SD$^{\textrm{a}}$  & 0.177	&	0.760	&	0.747	&	0.511	&	0.307	&	0.202 \\
Uncertainty$^{\textrm{b}}$  & 0.02 & 0.02 & 0.02 & 0.02 & 0.02 & 0.02\\
\hline
  Radial power (X-ray) \\
    $n$ value: & 0 & 1  & 2 & 3 & 4& 5 \\
   \hline
Mean & 1.130 &	1.064&	0.701&	0.671&	0.504&	0.376 \\
SD & 0.162  &	0.655&	0.676&	0.643&	0.622& 0.529 \\
\hline
 Radial power difference \\
  $n$ value: & 0 & 1  & 2 & 3 & 4& 5 \\
   \hline
   Mean &        -0.036&0.087&  0.181	&	0.200	&	0.212	& 0.192 \\
SD &0.178&0.672	&0.753	&	0.647	&0.598	&	0.517	\\	
\hline
Angular power  (radio) \\
    $m$ value: & 0 & 1  & 2 & 3 & 4& 5 \\
   \hline
Mean & 3.177	&	0.211	&	0.102	&	0.048	&	0.049	&	0.024  \\ 
SD & 2.300     &	0.184&	0.137&	0.101 &	0.075&	0.072  \\
\hline
Angular power  (X-ray) \\
    $m$ value: & 0 & 1  & 2 & 3 & 4& 5 \\
   \hline
Mean  & 4.034	&	0.223	&	0.103	&	0.036	&	0.035	&	0.016  \\ 
SD & 2.594     &	0.285&	0.143&	0.058 &	0.046&	0.034  \\
\hline
Angular power difference \\
 $m$ value: & 0 & 1  & 2 & 3 & 4& 5 \\
 \hline
Mean &	 0.857	&0.012	&	0.001		&-0.012	&	-0.014&	-0.008\\	
SD     & 2.591 &0.310&	0.127	&	0.073	&	0.073	&	0.043 \\
\hline
    \end{tabular}
    \label{tab:xrdiff}
\end{center}
\tablecomments{$^{\textrm{a}}$: SD means standard deviation.}
\tablecomments{$^{\textrm{b}}$: The calculation of uncertainty in individual radial and angular powers is described in Section~\ref{sec:SNRmult}. They are the same for different multipoles, thus this row is not repeated in the table for the different cases in Column 1.}
\end{table*}

Table~\ref{tab:xrdiff} lists the statistics (means and standard deviations) of the radial powers and the angular powers, derived from the radio images and from the X-ray images.
The standard devations (SDs) are shown to illustrate the diversity of multiple powers within any given subset.
The statistical errors in the multipole powers are in most cases much less than the SDs, with the exception being the $m$=4 and 5 angular powers, where the means are $\sim$1 to 3 times the noise.
The statistics of the  difference between radio and X-ray powers are also given. 

The mean radial powers decrease with increasing order for both X-ray and radio images.
The decrease is gradual from $n=0$ to 5: by a factor of $\simeq$6 for radio, and by a factor of $\simeq$3 for X-ray.

The mean angular powers also decrease with increasing order in both X-ray and radio. 
Angular powers show a much faster decrease with increasing order than radial powers. 
From $m=0$ to 5 the angular powers decrease by factor of $\simeq$130 for radio, and by a factor of $\simeq$250 for X-ray.

The difference between radio structure and X-ray structure of each SNR can be measured using the power difference (the X-ray power minus the radio power for each SNR).
Including both radial and angular cases gives 12 cases ($n=0$ to 5 and $m$=0 to 5).
The means and standard deviations of the differences are listed in Table~\ref{tab:xrdiff}.

There are differences larger than the noise between the radio and X-ray radial powers,  with the mean X-ray minus radio radial powers positive by more than the noise for $n$=1 through 5. 
The cases (of the 168) for which the individual radio and X-ray radial powers are the same within the noise are- 
for $n$=0: G315.4-2.3, G337.2 -0.7, G346.6-0.2, G11.2-0.3, G21.5-0.9, G27.4+0.0, G74.0-8.5, G292.0 +1.8; 
$n$=1: G33.6+0.1;
$n$=2: G111.7-2.1, G327.4 +0.4; 
$n$=3: G306.3 -0.9, G352.7-0.1, G33.6+0.1, G332.4-0.4;
$n$=4: G120.1 +1.4, G327.6 +14.6, G111.7 -2.1;
and $n$=5: G1.9 +0.3, G306.3 -0.9, G332.4-0.4.
I.e. only 21 of 168 radial powers are the same within noise comparing radio and X-ray, with 8 of these for $n$=0.
The SNRs with radio and X-ray radial powers consistent with each other are different for different values of $n$.

For the angular structure, the mean X-ray minus radio powers are less than the noise for $m$=1 through 5. 
There are many individual cases for which the angular radio and X-ray powers are the same within the noise- 
for $m$=0: 1 case;
$m$=1: 3 cases; 
$m$=2: 8 cases; 
$m$=3: 15 cases;
$m$=4:  15 cases; 
and $m$=5:  23 cases.  
I.e., 65 of 168 angular powers are the same within noise comparing radio and X-ray, significantly more than for the radial powers.

In summary, no SNR has all radio and X-ray powers consistent with each other, which implies all SNRs have some differences between their radio and X-ray structure.
All SNRs have most of their radial powers different between radio and X-ray. 
The mean radial powers are larger for X-ray than radio for $n$=1 through 5 and similar for $n$=0.
In contrast, approximately $2/3$ of SNRs have angular powers different between radio and X-ray.
The mean radio and X-ray angular powers are consistent within the noise for $m$=1 through 5.
Thus overall, there is evidence for larger X-ray than radio radial powers but similar X-ray and radio angular powers for SNRs.

\subsection{Comparison of Powers Between Type Ia and Type CC SNRs}

\begin{table*}[h!]
\begin{center}
\caption{Statistics of Power-Ratios for the Type Ia SNRs and Type CC SNRs}
   \centering
    \begin{tabular}{lccccccccc}
    \hline
 Sample: & Radio SNRs  &&X-ray SNRs \\ 
  &  11 Type Ia Radio & 17 Type CC Radio & 11 Type Ia X-ray & 17 Type CC X-ray \\
\hline
Mean $P_{rad,0}$ 	&1.207 &1.140	&1.158	&1.112	\\
SD$^{\textrm{a}}$ $P_{rad,0}$     &0.224 &0.131 &0.216 &0.109 \\
Uncertainty$^{\textrm{b}}$ & 0.02 & 0.02 & 0.02 & 0.02 \\
Mean $P_{rad,1}$ &0.785 	&1.101	&0.782 &1.247			\\
SD $P_{rad,1}$&0.510 &   0.862 &0.432 &0.707	\\
Mean $P_{rad,2}$ &0.298 &0.663	&0.427 &0.878	\\
SD $P_{rad,2}$ &0.209 &0.916&	0.331 &0.776		\\
Mean $P_{rad,3}$ &0.315 &0.572	&0.461	&0.807			\\
SD $P_{rad,3}$ & 0.238 &  0.606 &0.327 &0.751	\\
Mean $P_{rad,4}$ &0.239 &0.327 &0.348	&0.604		\\
SD $P_{rad,4}$ &0.152 &0.371 &0.343	&0.732\\
Mean $P_{rad,5}$ &0.133	&0.219	&0.237	&0.467	\\
SD $P_{rad,5}$&0.102 &0.240 &0.248 &	0.633\\
\hline
Mean $P_{ang,0}$& 2.473& 3.633 &3.001 &4.703	\\		
SD $P_{ang,0}$ &0.922& 2.763 &1.603 &2.879\\
Mean $P_{ang,1}$& 0.217 & 0.208 &0.165&0.261	\\
SD $P_{ang,1}$& 0.204& 0.169&  0.213 &0.317 \\
Mean $P_{ang,2}$ & 0.124& 0.088 &0.135	&0.083\\
SD $P_{ang,2}$&0.189 & 0.084  &0.197 & 0.088 \\
Mean $P_{ang,3}$& 0.076 & 0.031 &0.043 &0.031 		\\
SD $P_{ang,3}$& 0.151 & 0.036& 0.071 &	0.048\\
Mean $P_{ang,4}$ &0.045 &0.051 &0.043	&0.030 	\\
SD $P_{ang,4}$&0.051 & 0.087 &0.064&	0.029 \\
Mean $P_{ang,5}$ & 0.043&0.011&0.025 &0.009		\\
SD $P_{ang,05}$& 0.111& 0.015&	0.052 &0.010 \\
\hline
\hline
    \end{tabular}
    \label{tab:IavsCC}
\end{center}
\tablecomments{$^{\textrm{a}}$: SD stands for standard deviation.}
\tablecomments{$^{\textrm{b}}$: The calculation of uncertainty in radial and angular power values is described in Section~\ref{sec:SNRmult}. They are the same for different multipole powers (different radial $n$ or angular $m$ values), thus this row is not repeated in the table for the different cases in Column 1.}
\end{table*}

Table~\ref{tab:IavsCC} lists means and standard deviations for the radial and angular powers derived from the radio and X-ray images of the SNRs separately for Type Ia and Type CC SNRs.
For all modes (12 cases of $n$ and $m$) and for all four samples (Radio Type Ia, Radio Type CC, X-ray Type Ia, and X-ray Type CC) the SDs are significantly larger than the individual errors in powers of $\sim0.02$.
Thus there are differences between the different SNRs within each of the four samples.

\subsubsection{Comparison of Powers Between Type Ia and Type CC SNRs in Radio}

First we compare the mean powers from the radio images between Type Ia and Type CC SNRs.
The mean radial powers decrease with increasing $n$ for both Type Ia and CC, with the single exception that the $n=3$ power is larger than the $n=2$ power for Type Ia. 
The decrease with $n$ is more rapid for Type Ia (factor of $\sim$9) than for CC (factor of $\sim$5).
For all radial modes ($n$) the difference between the means for Type Ia and Type CC is smaller than their SDs.
Except for $n=0$, the powers are larger for Type CC SNRs than for Type Ia SNRs.

The angular powers also decrease with increasing $m$ for both Type Ia and CC, with the single exception that the $m=4$ power is larger than the $m=3$ power for Type CC. 
In contrast to radial powers, the decrease with $m$ is more rapid for Type CC (factor of $\sim$300) than for Type Ia (factor of $\sim$60).
For all angular modes ($m$) the difference between the means for Type Ia and Type CC is smaller than their SDs.

\subsubsection{Comparison of Powers Between Type Ia and Type CC SNRs in X-ray}

The mean X-ray radial powers decrease with increasing $n$ for both Type Ia and CC, with the two exceptions that the $n=3$ power is larger than the $n=2$ power for Type Ia, and the $n=1$ power is larger than the $n=0$ power for Type CC. 
The decrease with $n$ is more rapid for Type Ia (factor of $\sim$5) than for CC (factor of $\sim$2.4).
For all radial modes ($n$) the difference between the means for Type Ia and Type CC is smaller than their SDs.
Except for $n=0$, the powers are larger for the Type CC SNRs than for Type Ia SNRs.

The angular powers also decrease with increasing $m$ for both Type Ia and CC. 
In contrast to the radial powers, the decrease with $m$ is more rapid for Type CC  (factor of $\sim$500) than for Type Ia (factor of $\sim$120).
For all angular modes ($m$) the difference between the means for Type Ia and Type CC is smaller than their SDs.

In summary, both radio and X-ray images show similar trends. 
The radial powers are larger for Type CC than Type Ia SNRs and the angular powers are similar between Type CC and Type Ia SNRs. 
The radial powers in X-ray are larger than those in radio for $n$=1 to 5 in radio for both Type CC and Type Ia. 
The angular powers in X-ray are similar to those in radio for most cases of $m$ for both Type CC and Type Ia. 

\subsection{Kolmogorov-Smirnov (KS) Test Applied to Cumulative Probability Distribution Functions (CDFs)}

\begin{table}[h!]
\begin{center}
\caption{Threshold differences $D_{n,m}$ for different confidence levels for the Kolmogorov-Smirnov test}
   \centering
    \begin{tabular}{lccccc}
    \hline
 CL$^a$: & 80\%  & 90\% & 95\% & 99\%  &99.5\%  \\ 
 \hline
(i,j)$^b$ & &  & &\\
\hline
(17,11) 	&0.415 & 0.474	& 0.526	& 0.630 & 0.670	\\
(28,28)&0.287 & 0.327	& 0.363	& 0.435 & 0.463	\\
(13,11)&0.440& 0.501	& 0.556	& 0.667& 0.709	\\
\hline
    \end{tabular}
    \label{tab:KS}
\end{center}
\tablecomments{$^{\textrm{a}}$: CL means confidence level.}
\tablecomments{$^{\textrm{b}}$: i and j are the sample sizes for the two CDF's.}
\end{table}

We apply the Kolmogorov-Smirnov (KS) test to see if the distributions of powers for Type Ia SNRs and Type CC SNRs are statistically different. 
The KS test is applied to the cumulative probability distribution functions (CDF) of two variables. 
It yields the confidence level (CL) of the difference from the maximum difference in the CDFs.
Table~\ref{tab:KS} shows the threshold difference, $D_{n,m}$, for two cases: $(i,j)=(17,11)$ relevant to comparing the Type CC and Ia samples, and $(i,j)=(28,28)$ relevant to comparing various quantities for the whole SNR sample of 28.
E.g. for $(i,j)=(17,11)$ for the Type Ia and Type CC samples, 90\% confidence that two CDFs are different (not consistent with each other) corresponds to $D_{n,m}$=0.474.
As the sample sizes increase, the difference in CDFs required to conclude (at a given confidence level) that the two CDFs are different decreases.
I.e., larger samples mean that it is easier to distinguish PDFs.

\subsubsection{KS Test to Compare Type Ia and Type CC Powers}

\begin{figure*}
\centering
\includegraphics[width=0.45\linewidth]{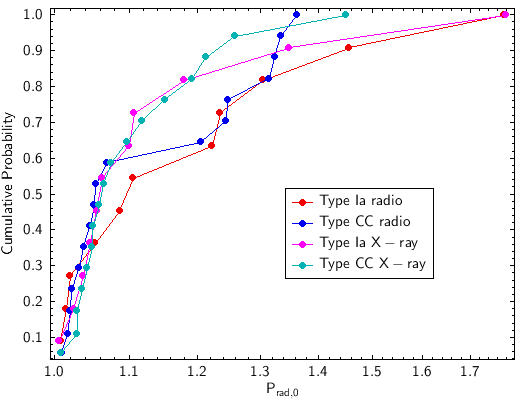}
\includegraphics[width=0.45\linewidth]{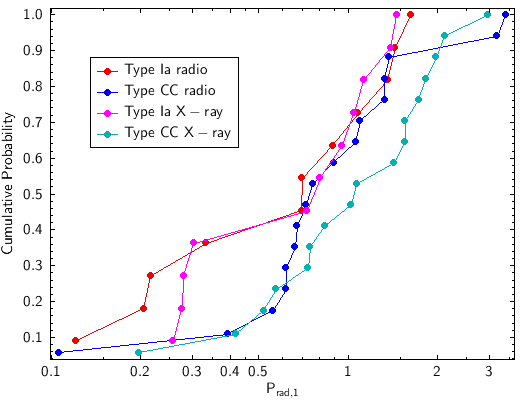}
\includegraphics[width=0.45\linewidth]{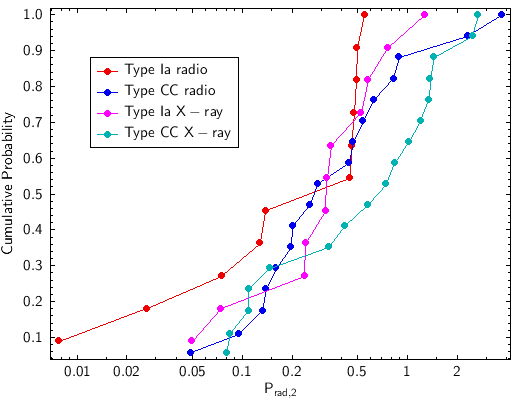}
\includegraphics[width=0.45\linewidth]{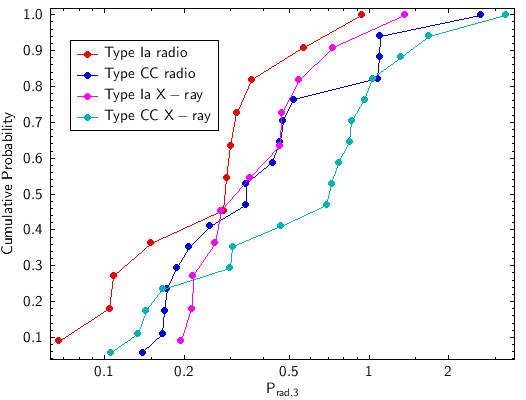}
\includegraphics[width=0.45\linewidth]{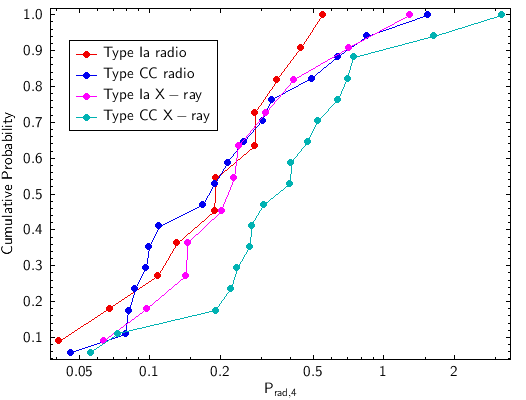}
\includegraphics[width=0.45\linewidth]{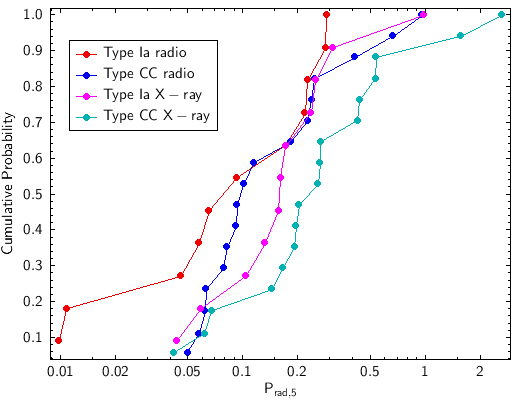}
\caption{CDF (cumulative probability distribution functions) comparing Type Ia and Type CC radial powers for radio images (red and blue) and X-ray images (magenta and cyan); for $n=0$ to 5 (top left to bottom).
}
\label{fig:CDFrad}
\end{figure*}

\begin{figure*}
\centering
\includegraphics[width=0.45\linewidth]{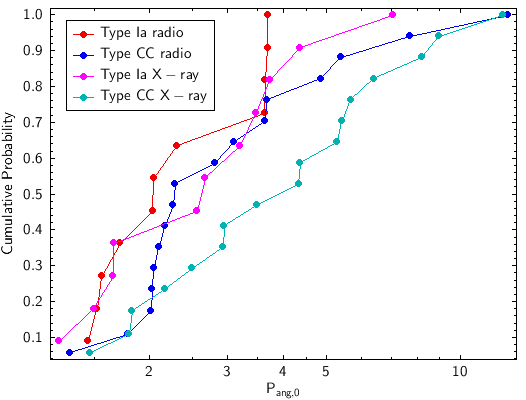}
\includegraphics[width=0.45\linewidth]{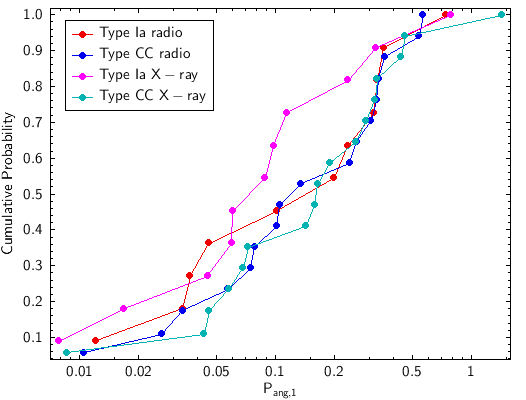}
\includegraphics[width=0.45\linewidth]{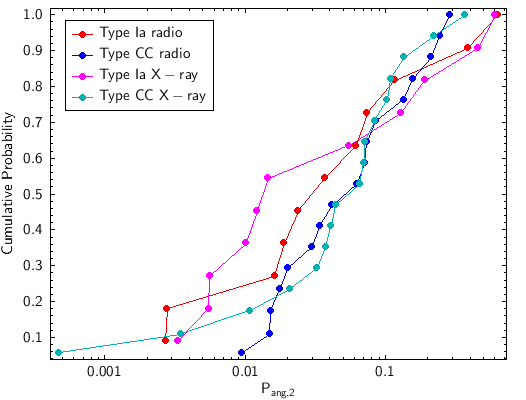}
\includegraphics[width=0.45\linewidth]{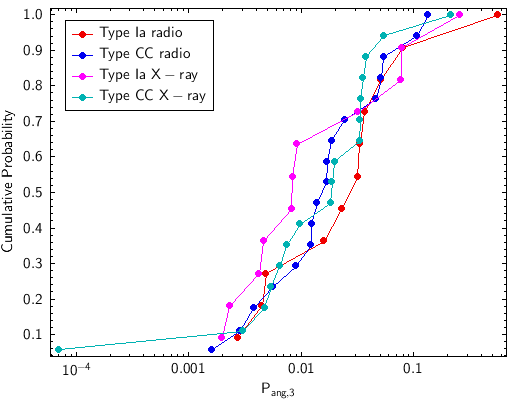}
\includegraphics[width=0.45\linewidth]{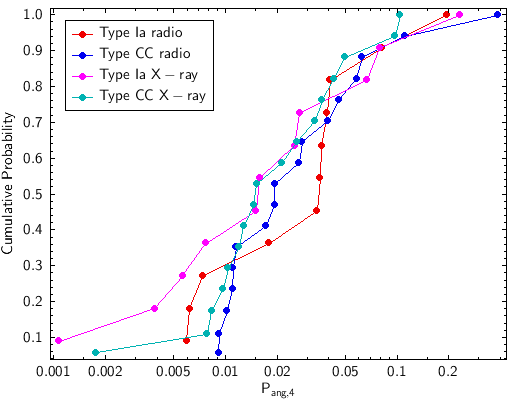}
\includegraphics[width=0.45\linewidth]{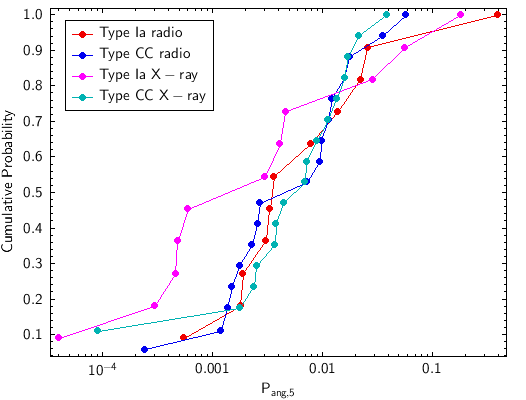}
\caption{CDF (cumulative probability distribution functions) comparing Type Ia and Type CC angular powers for radio images (red and blue) and X-ray images (magenta and cyan); for $m=0$ to 5 (top left to bottom).
}
\label{fig:CDFang}
\end{figure*}

The CDFs of the radial powers for Type Ia and Type CC SNRs are shown in Figure~\ref{fig:CDFrad}, calculated from radio and X-ray images. 
The Type Ia CDFs (red) are compared with Type CC CDFs (blue) for radio, and the Type Ia (magenta) with Type CC (cyan) CDFs for X-ray.
A 90\% confidence difference corresponds to $D_{n,m}$=0.474.
None of the radial powers are significantly different at 90\% confidence between Types Ia and CC. 
The largest difference between Types Ia and CC CDFs is for $n=3$ for the X-ray images at 0.43 which is $\sim$83\% confidence. 
A larger sample size (smaller $D_{n,m}$) would help to conclude significant differences between Type Ia and Type CC.
Although the CDFs are not significantly different by the KS test, the Type CC X-ray CDFs show more radial powers (displaced to the right) than the other three sets for all $n>$0.

The CDFs of the angular powers are shown in Figure~\ref{fig:CDFang}, calculated from the radio images and from the X-ray images. 
The largest CDF difference of 0.34 is for the $m$=5 case for Type Ia and Type CC X-ray powers.
None of the angular powers are significantly different at the 90\% level.

\subsubsection{KS Test to Compare Radio and X-ray Powers of SNRs}

\begin{figure*}
\centering
\includegraphics[width=0.45\linewidth]{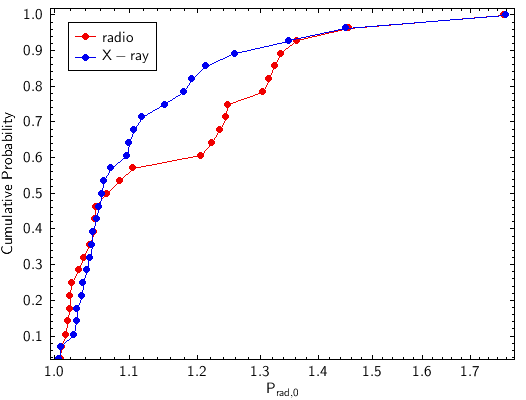}
\includegraphics[width=0.45\linewidth]{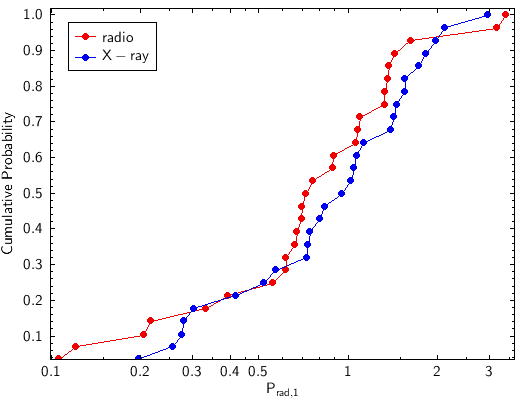}
\includegraphics[width=0.45\linewidth]{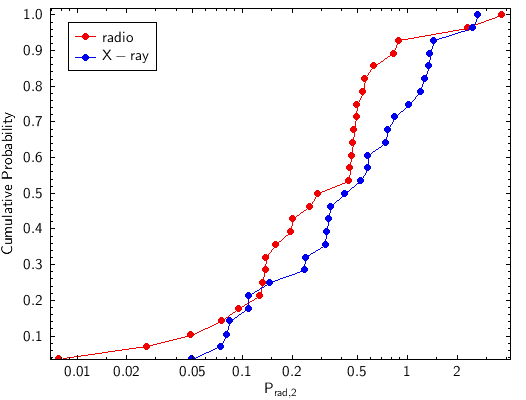}
\includegraphics[width=0.45\linewidth]{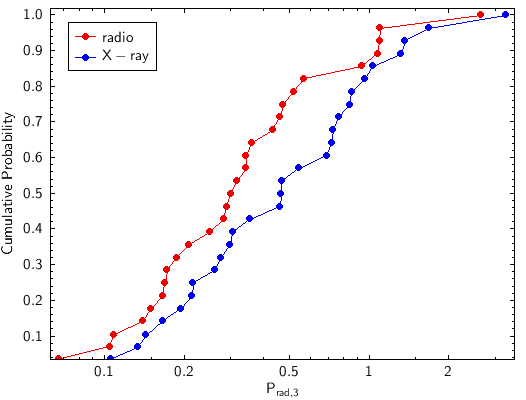}
\includegraphics[width=0.45\linewidth]{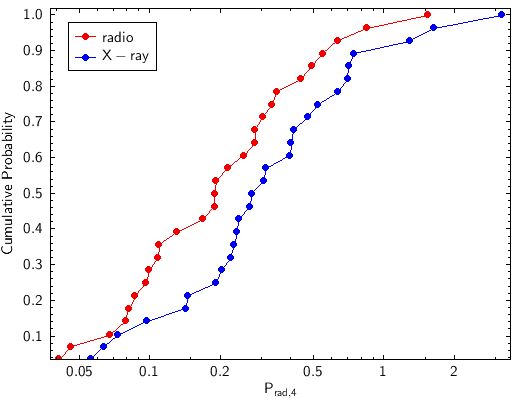}
\includegraphics[width=0.45\linewidth]{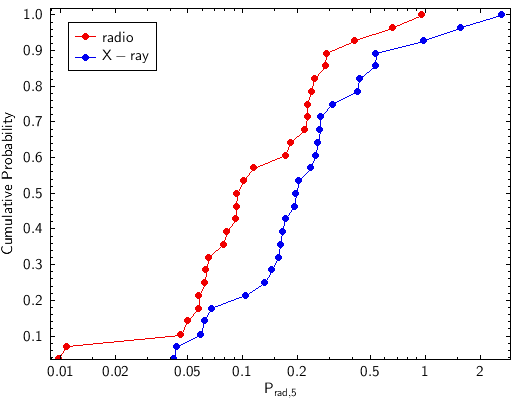}
\caption{CDFs (cumulative probability distribution functions) compared for radio images and X-ray images. Radial power CDFs for $n=0$, 1, 2, 3, 4 and 5 are shown. 
}
\label{fig:CDF1}
\end{figure*}

\begin{figure*}
\centering
\includegraphics[width=0.45\linewidth]{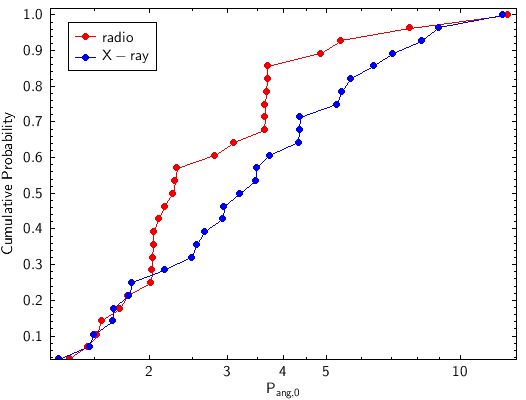}
\includegraphics[width=0.45\linewidth]{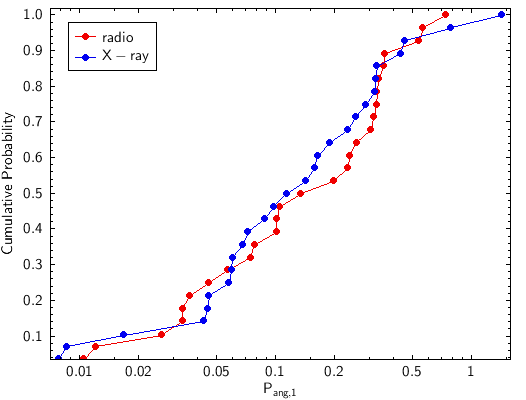}
\includegraphics[width=0.45\linewidth]{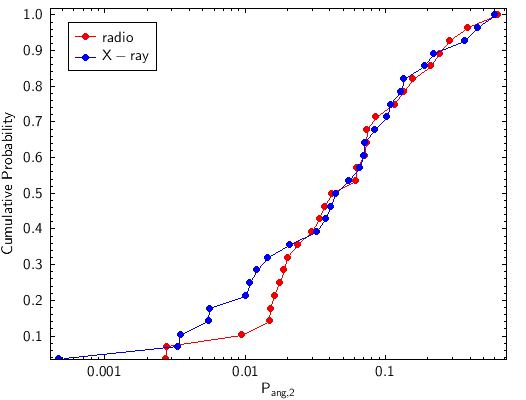}
\includegraphics[width=0.45\linewidth]{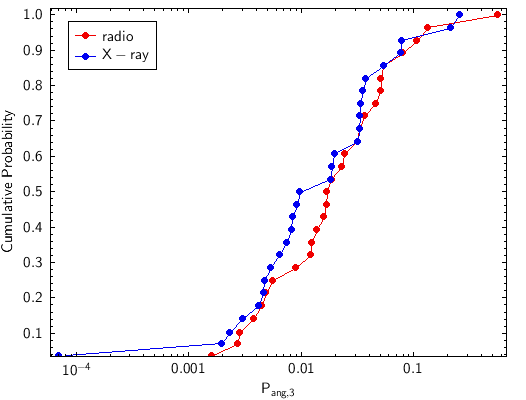}
\includegraphics[width=0.45\linewidth]{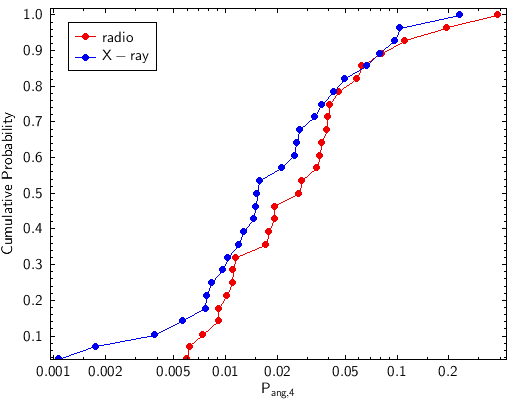}
\includegraphics[width=0.45\linewidth]{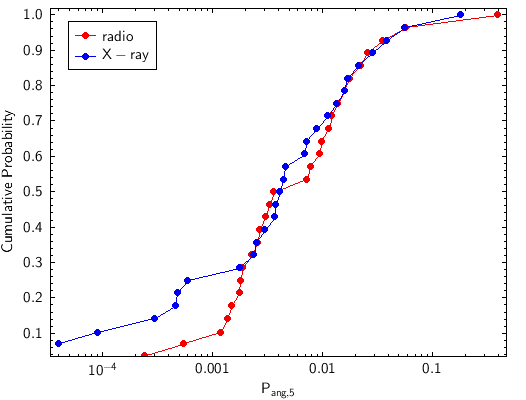}
\caption{CDFs (cumulative probability distribution functions) compared for radio images and X-ray images. Angular power CDFs for $m=0$, 1, 2, 3, 4 and 5 are shown.
}
\label{fig:CDF2}
\end{figure*}

For this sample (28 radio images and 28 X-ray images), a 90\% confidence difference corresponds to $D_{n,m}$=0.363, and 99\% to $D_{n,m}$=0.435.
The CDFs of the radial powers for radio  and X-ray are shown in Figure~\ref{fig:CDF1}, calculated from the radio images and from the X-ray images.
The largest CDF difference of 0.35 is for the $n$=5 case.
None of the CDFs for the radial powers reach the 90\% confidence difference between radio and X-ray.
For $n$=0 the radio powers are larger (CDF displaced to larger x-axis values), but for $n$=1, 2, 3, 4 and 5 the X-ray powers are larger.

The CDFs of the angular powers are shown in Figure~\ref{fig:CDF2}. 
None of the CDFs for the angular powers reaches the 90\% confidence difference between radio and X-ray.
The $m$=0 mode has more power in X-ray compared to radio, whereas the $m$=1, 2, 3, 4 and 5 modes have very similar powers in radio and X-ray.  

In summary, the KS test does not yield a significant difference at 90\% level between the CDFs of radio and X-ray powers.
However, the CDFs show a clear displacement for radial modes ($n$=1 to 5) with X-ray powers larger than radio multipole powers, but no difference for angular modes ($n$=1 to 5) between X-ray and radio multipole powers.

\begin{figure*}
\centering
\includegraphics[width=0.49\linewidth]{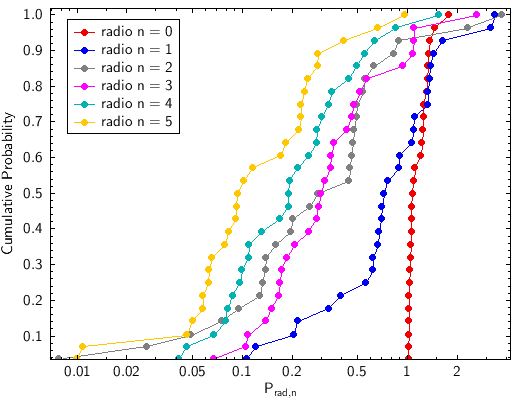}
\includegraphics[width=0.49\linewidth]{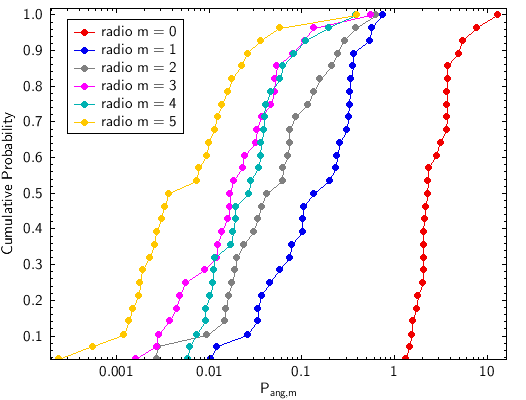}
\includegraphics[width=0.49\linewidth]{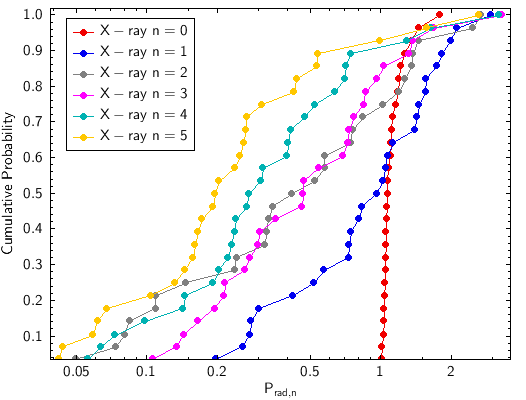}
\includegraphics[width=0.49\linewidth]{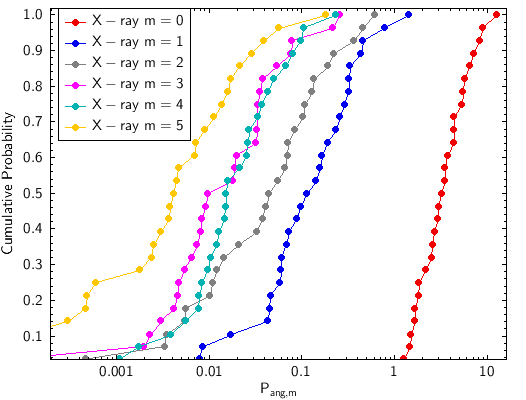}
\caption{CDFs (cumulative probability distribution functions) of different radial modes and different angular modes compared. 
The radial modes on the left and the angular modes are on the right, with values for radio images on the top and values for X-ray images on the bottom row.
}
\label{fig:CDF3}
\end{figure*}

Next the KS test is applied to test for significant differences between different modes.
Figure~\ref{fig:CDF3} shows the CDFs of radial modes( $n$=0 to 5) compared and different angular modes ($m$=0 to 5) compared. 
Radio CDFs (top two panels) and X-ray CDFs (bottom two panels) are shown. 

For radial modes of radio images (top left panel), the differences exceed 99\% confidence (difference $>$0.435) for most cases (except comparing $n$=2 with 3; 2 with 4; 3 with 4; and 4 with 5).
For radial modes of X-ray images (bottom left panel), the differences exceed 99\% confidence only for $n$=0 compared with any other $n$ and $n$=1 with 4 or 5.
For angular modes for radio images (top right panel), the differences exceed 99\% confidence for most cases (except comparing $n$=1 with 2; 2 with 3 or 4; and 3 with 4 or 5).
For angular modes for X-ray images (bottom right panel), the differences exceed 99\% confidence for most cases (except comparing $n$=1 with 2; 2 with 3 or 4; 3 with 4 or 5; and 4 with 5).

The powers generally decrease with increasing $n$ or $m$, with a few exceptions where adjacent modes have similar powers: $n$=2 and 3 for radio radial powers; $n$=3 and 4 for radio angular powers; $n$=2 and 3 for X-ray radial powers; and $n$=3 and 4 for X-ray angular powers.
Thus the different modes have significantly power distributions for most cases, with the cases for less than 99\% confidence in difference usually being adjacent $n$ or $m$ values. 

\section{Discussion}\label{sec:disc}

\subsection{Results of the current work}\label{sec:discNew}

We derived new basis functions for multipole analysis which are equal area weighted. 
The unweighted radial basis functions derived by \cite{2025PASP..137f4502L} were used because they were straight-forward to derive using the Gramm-Schmitt normalization method. 
That gave equal r and $\phi$ coordinate weighting ($dr\times d\phi$), which because an area element is $dr\times rd\phi$ gives more weight to smaller radii compared to equal area weighting.
Equal area weighting is the most commonly used method for most cases of astronomical image analysis.

The multipole analysis was applied to a larger sample of SNR images than previously considered. 
Both radio and X-ray images are analyzed, which has not been done before.

The analysis yields the following results: 
\begin{itemize}
\item The multipole radial powers from $n=0$ through $n=5$ and
angular powers from $m=0$ through $m=5$ are different comparing one SNR to another. This is not surprising because the structures of SNRs are different from each other by manual inspection.
\item The multipole powers computed from the X-ray image of an SNR are different than the multipole powers from the radio image of that same SNR. This quantifies the difference between an SNR's X-ray and radio images. 
\item The averages and standard deviations of the multipole powers for radio images and for X-ray images, as well as the power differences (radio power minus X-ray power of each SNR) are summarized in Table~\ref{tab:xrdiff}. The standard deviations are much larger than the noise level for all powers (for all modes, for X-ray powers, for radio powers and for their difference). 
\item The cumulative probability distributions (CDFs) of powers are given in Figure~\ref{fig:CDF1} for radial powers and Figure~\ref{fig:CDF2} for angular powers. 
X-ray powers are higher than radio powers for radial modes $n$=1, 2, 3, 4 and 5, but for angular powers X-ray and radio images have similar CDFs for $m$=1, 2, 3, 4 and 5.
None of the modes have CDFs for which the KS test yields difference between X-ray and radio at $>$90\% confidence.
\item The mean powers for Type Ia SNRs and Type CC SNRs are different by more than the noise level, but smaller than the standard deviation within each group. 
Radial powers for $n$=1 to 5 are larger for Type CC than Type Ia for X-ray and radio with larger difference in X-ray than radio. The angular powers (Figure~\ref{fig:CDFang})  for $n$=1 to 5 show no clear difference between Types Ia and CC, either for X-ray or radio. 
\end{itemize}

Figure~\ref{fig:CDF3} shows a summary of the CDFs of $P_{rad,n}$ and $P_{ang,m}$ for the radio images and for the X-ray images. 
Generally, there is a trend of decreasing powers as $n$ or $m$ increases, with a few exceptions. 
E.g. $P_{rad,n}$ for $n$=2 and 3 are similar (radio and X-ray), and $P_{ang,m}$ for $m$=3 and 4 are similar.
Overall, we find most SNRs have more power (more structure) in their X-ray image than in their radio image. 

\subsection{Comparison with Previous Work}

A multipole analysis was applied to X-ray images of SNRs by \cite{2011ApJ...732..114L}, using an incomplete and non-orthogonal set of basis functions. 
That sample included 24 X-ray images of SNRs.
They found no difference between the two types when applying a wavelet analysis, but did find a difference using their incomplete and non-orthogonal basis functions. 
The same non-orthogonal and incomplete basis functions were applied to radio images of SNRs by \cite{2019arXiv190911803R} who found no distinction between Type Ia and Type CC SNRs. 

A more recent study using multipole analysis with complete and orthogonal basis functions was carried out by \cite{2025PASP..137f4502L}.
That work analyzed the same 24 X-ray images as \cite{2011ApJ...732..114L}, and found no significant population difference between Type Ia and Type CC SNRs 
using the image center as center for analysis. 
That work also compared PRM and complete/orthogonal basis function multipoles when the centroid was used as center for analysis (their Fig. 6). 
They did not apply the KS test, so we applied the KS test to the PRM and complete/orthogonal multipole powers of their Fig.6, i.e. PRM powers $P_2/P_0$ and  $P_3/P_0$ and  complete/orthogonal  powers $P_{2,2}$ and  $P_{3,3}$. 
The cumulative probability distributions give maximum differences in cumulative probability between the sample of 11 Type Ia SNRs and sample of 13 Type CC SNRs of 0.552, 0.231, 0.292 and 0.290, respectively, for the 4 powers above.
The KS threshold values for different confidence values for samples sizes of 11 and 13 are shown in Table~\ref{tab:KS} above.
This shows that the difference between Type Ia and Type CC is not significant for the powers  $P_3/P_0$, $P_{2,2}$ and  $P_{3,3}$, and that for $P_2/P_0$  it reaches above 90\% but below 95\% confidence. 
The real significance is lower, because we have applied that KS test multiple times and, e.g., for 20 applications of the KS test one would expect one to reach 95\% confidence just by chance.

The current work improves on previous studies in the following ways: new radial basis functions are derived here which are equal-area weighted; the analysis is extended to include multipole orders $n$ and $m$ from 0 to 5; the sample is larger, with 28 SNRs; and both X-ray and radio images are analyzed and compared.
Here we found more radial power in Type CC SNRs than for Type Ia SNRs, with the difference larger for X-ray images than for radio images. 
We also found none of the CDFs of a given multipole power
have significant differences exceeding 90\% confidence using the KS test.
The results here are generally consistent with previous studies, with indications of differences between X-ray and radio multipole powers and indications that Type CC SNRs have more power than Type Ia SNRs.
However the differences between individual SNR's structures dominates over population level differences with the current numbers of SNRs available for comparison. 
 
\section{Conclusions}

In the current study, we derive new basis functions for 2 dimensional (2D) image analysis (referred to as multipole analysis) with a circular boundary.  
The new basis functions are equal-area weighted, which is more natural for astronomical image analysis than other weighting methods.
The basis functions are complete and orthogonal: this is required for fully representing 2D images. 

The analysis is applied to a sample of 28 SNRs with both X-ray and radio images (total of 56 images), and which have been classified as Type Ia or Type CC.
A summary of the results of the multipole analysis follows.
The radial powers and angular powers, from order 0 to 5, for a given SNR are generally all different by more than the noise, and different between X-ray and radio images.
At a population level, there is a difference between X-ray and radio images, with the X-ray radial powers ($n>0$) larger than the radio ones (more structure in X-rays than radio). 
The angular powers are also larger in X-ray than in radio (see Figure~\ref{fig:CDF3}), but the angular powers in both X-ray and radio are smaller than the corresponding radial powers. 
Thus generally, SNRs show more radial structure than angular structure.
We find radial powers ($n>0$) larger for Type CC than Type Ia, with larger difference for X-ray images than for radio images. 
Angular powers ($m>0$) are similar for Type CC and Type Ia, for both radio and X-ray images.

When larger samples of SNR images are available in X-ray and radio in the coming years, more differences, and with higher significance, may be revealed between X-ray and radio structure, between radial and angular structure and between Type Ia and Type CC SNRs. 

{\large \noindent Acknowledgements}
This work supported by a grant from the Natural Sciences and Engineering Research Council of Canada.

\end{document}